\documentclass[%
 reprint,
 amsmath,amssymb,
 aps
]{revtex4-2}

\usepackage{graphicx}
\usepackage{dcolumn}
\usepackage{bm}
\usepackage[colorlinks=true,citecolor=blue]{hyperref}
\usepackage{url}

\begin{document}

\preprint{APS/123-QED}

\title{Study of topological quantities of lattice QCD with a modified Wasserstein generative adversarial network}

\author{Lin Gao$^1$}
\email{silvester\_gao@qq.com}

\author{Heping Ying$^2$}

\author{Jianbo Zhang$^2$}

\affiliation{\vspace{0.3cm}
\vspace{0.1cm} $^1$ American Association for the Advancement of Science, Washington, DC 20005, USA\\
\vspace{0.1cm} $^2$ School of Physics, Zhejiang University, Hangzhou, 310027, China}


\begin{abstract}
We propose a modified Wasserstein generative adversarial network (M-WGAN) to study the distribution of the topological charge in lattice QCD based on Monte Carlo simulations. We construct new generator and discriminator in M-WGAN to support the generation of high-quality distribution. Our results show that the M-WGAN scheme of machine learning should be helpful for us to calculate efficiently the 1D distribution of topological charge compared with the method by the MC simulation alone.
\end{abstract}

\maketitle


\section{Introduction}
Machine learning (ML) has improved the state-of-the-art in many domains\cite{Carrasquilla2017}\cite{GoodfellowNIPS2014}\cite{goodfellow2020generative}.Especially it has had many applications in lattice QCD. Kim A. Nicoli et al. exerted the deep generative models to estimate the absolute value of the free energy in lattice field theories\cite{Nicoli2021}. Ankur Singha et al. developed a conditional normalizing flow-based sampling method for scalar lattice $\phi^4$ theory to improve the problem of critical slowing down\cite{Singha2023}. Yuki Nagai et al. developed the self-learning Monte Carlo (MC) to resolve the autocorrelation problem\cite{Nagai2023}. Matteo Favoni and Andreas Ipp et al. propose Lattice gauge equivariant Convolutional Neural Networks to study applications of generic machine learning in lattice gauge theory\cite{Favoni2022}. These works have shown the great impact of ML in lattice QCD, including the calculation of physical quantities, the improvement of research methods, the reduction in autocorrelation and the study of gauge invariant quantities, etc. The conventional computation of topological charge in lattice QCD based on the MC method consumes a lot of computation time\cite{Wilson1974PRD}\cite{Gattringer2010}, and ML could be explored as a potential method for generating topological quantities efficiently. Our research focuses on constructing new generator and discriminator of Wasserstein generative adversarial network (M-WGAN) to facilitate the exploration of the topological quantities in lattice QCD. It is committed to generating corresponding topological charge data based on a small amount of MC data to calculate the topological susceptibility. In the following, we begin our studies with a brief introduction to the lattice QCD and ML.

In QCD, it is difficult to use perturbation methods like quantum electrodynamics in the low-energy region since the coupling constant of QCD becomes large\cite{GeorgiPolitzer1976}. Therefore, lattice QCD was introduced to study the QCD in the low-energy region\cite{Wilson1974PRD}, where Euclidean spacetime instead of Minkowski spacetime is used and this simplifies the computations as real numbers rather than complex numbers are involved. Furthermore, the action used here is the Wilson gauge action\cite{Gattringer2010}
\begin{equation}
{S_G}\left[ U \right] = \frac{\beta }{3}\mathop \sum \limits_{n \in {\rm{\Lambda }}} \mathop \sum \limits_{\mu  < \nu} {\mathop{\rm Re}\nolimits} {\rm{Tr}}\left[ {1 - {U_{\mu \nu}}\left( n \right)} \right],
\end{equation}
where $\beta$ is the inverse coupling and $U_{\mu \nu}\left(n\right)$ is the plaquette. 

The topological charge density discussed in this article is defined by 
\begin{equation}
q\left(n\right)=\frac{1}{32\pi^2}\varepsilon_{\mu \nu\rho\sigma}{\rm{Re}}{{\rm{Tr}}{\left[F_{\mu\nu}^{clov}\left(n\right)F_{\rho\sigma}^{clov}\left(n\right)\right]}},
\end{equation}
related to the non-conserved axial vector current\cite{tHooft1986}. $F_{\mu\nu}^{clov}\left(n\right)$ is the clover improved lattice discretization of the field strength tensor $ F_{\mu\nu}\left(x\right)$ and can be noted as
\begin{equation}
\begin{split}
    F_{\mu\nu}^{\text{clov}}(n) = & -\frac{i}{8a^2} \biggl[ \left( C_{\mu\nu}(n) - C_{\mu\nu}^\dag(n) \right) \\
    & - \frac{\text{Tr} \left( C_{\mu\nu}(n) - C_{\mu\nu}^\dag(n) \right)}{3} \biggr],
\end{split}
\end{equation}
where the clover is given as 
\begin{equation}
C_{\mu\nu}(n)=U_{\mu,\nu}(n)+U_{\nu,-\mu}(n)+U_{-\mu,-\nu}(n)+U_{-\nu,\mu}(n).
\end{equation}
The topological charge is further introduced as
\begin{equation}
Q_{top}=a^4\sum_{n\in\mathrm{\Lambda}} q\left(n\right),
\end{equation}
which converges to an integer in the continuum limit\cite{atiyah1971index}. The above non-conserved axial vector current is related to a symmetry breaking which is called ${U\left(1\right)}_A$ anomaly.From a physical point of view, a possible explanation is that instanton can be introduced to obtain the conserved charge\cite{BelavinPolyakov1975}. Furthermore, the topological susceptibility $\chi_t$ is defined as
\begin{equation}
\chi_t\ =\frac{1}{V}\left\langle{Q_{top}}^2\right\rangle,
\end{equation}
where $V$ is the 4D volume. The $\chi_t$ is related to the Witten-Veneziano relation which indicates that $\chi_t$ is proportional to the mass squared of $\eta^\prime$ meson for massless quarks\cite{witten1979current}. In our investigations, when computing the topological susceptibility from the initial configurations, we incorporate the Wilson flow as a technique to smooth the configurations and alleviate ultraviolet (UV) divergences.\cite{reuter1997renormalization}\cite{Luscher2010}\cite{Zhang2010rn}.

On the other hand, ML can be classified into supervised learning, unsupervised learning, semi-supervised learning, transfer learning and reinforcement learning. Supervised learning in machine learning typically demands millions of training examples to attain optimal results\cite{goodfellow2020generative}. However, it proves impractical when dealing with scenarios involving just a few hundred training data points. Generative adversarial networks (GANs) were proposed in 2014\cite{GoodfellowNIPS2014}, which would gradually improve the capabilities of generator and discriminator through their mutual adversarial game. It is often necessary to know the probability distribution of a set of data, such as the age distribution of a biological population or the distribution of pixels in an image. Compared with the method of maximum likelihood estimation which directly estimates the parameters of the probability density, GANs are implicit models that can infer the probability distribution p(x) without explicitly expressing the probability density function\cite{goodfellow2020generative}. For simple distribution like the age distribution of a biological population, one can guess a probability density function and then use the data to estimate the parameters of the probability density function. However, the probability density function may have millions or even billions of parameters for complicated distribution, and it will be difficult to guess a probability density function in this case. In contrast, the implicit models GANs have a good performance in data generation for the complex distribution. The topological charge that we are going to discuss obeys a certain distribution, so GANs can be used to generate data of the topological charge. After GANs were proposed, there have been many variants, one of which is Wasserstein generative adversarial network (WGAN)\cite{arjovsky2017WGAN}. The WGAN introduces the Wasserstein distance
\begin{equation}
\mathbb{E}_{x\sim\mathbb{P}_{r}}[D(x)]-\mathbb{E}_{x\sim\mathbb{P}_{G}}[D(x)] ,
\end{equation}
where $\mathbb{P}_r$ is the distribution of real data, $\mathbb{P}_G$ is the distribution generated by the generator and the discriminator $D(x)$ is required to be a 1-Lipschitz function. As a result, WGAN greatly improves the stability of GAN training and the quality of results.

\section{Model and data preparations}
For our purpose, the suitable ML model should be constructed to explore the characteristics of the topological charge in lattice QCD. We checked that the results will be poor if the neural networks of generator and discriminator are too simple in WGAN through testing. For example, if we use a simple neural network to generate the distribution of the topological charge, we will find that the symmetry of the distribution is not good or many bars of the distribution histogram stick together instead of being discrete.Therefore, the new generator and discriminator are constructed to study the distribution of the topological charge. 

\begin{figure}[htb]
\includegraphics[width=0.4\textwidth]{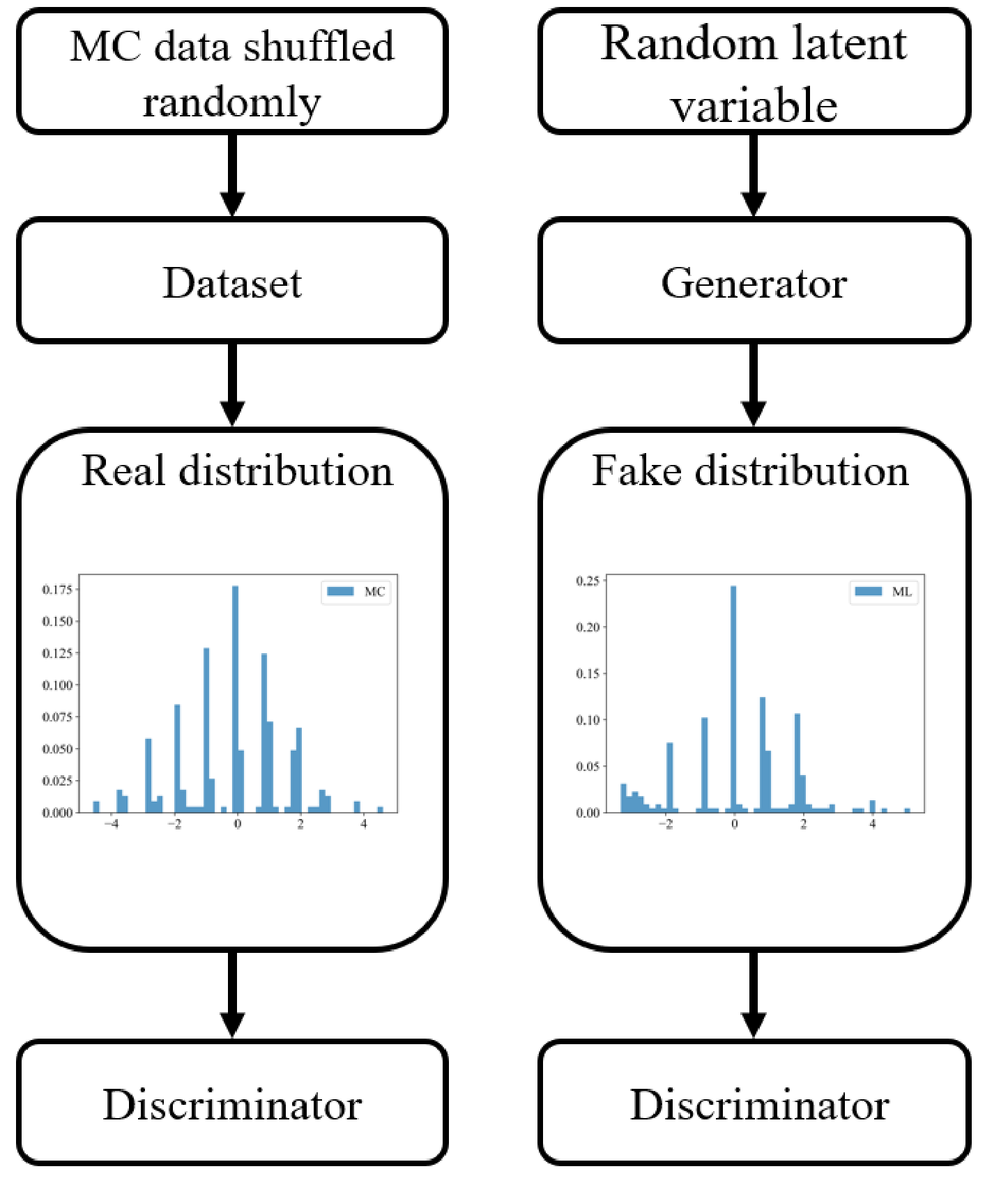}
\caption{\label{figure_1_ref}The overall structure of M-WGAN.}
\end{figure}

The overall structure of M-WGAN is shown in Fig.~\ref{figure_1_ref}.The real distribution in Fig.~\ref{figure_1_ref} refers to the distribution generated by the MC method, and the fake distribution refers to the distribution generated by machine learning at a certain epoch in the training process. These two distribution histograms are only used to help illustrate the outline of M-WGAN in the training process.The training process of M-WGAN is similar to that of WGAN\cite{arjovsky2017WGAN}\cite{goodfellow2020generative}. The differences are mainly in the sampling method of training data and the structure of the generator and the discriminator.

\begin{table}[htb]
\caption{\label{generator_structure_tab}The structure of the generator for M-WGAN.}
\begin{ruledtabular}
\begin{tabular}{llr}
\textrm{Layer (type) }&
\textrm{Output Shape }&
\textrm{Parameter number }\\
\colrule
Linear-1&[-1, 6400]&6,400\\
BatchNorm1d-2&[-1, 6400]&12,800\\
LeakyReLU-3&[-1, 6400]&0\\
Linear-4&[-1, 100]&640,000\\
BatchNorm1d-5&[-1, 100]&200\\
LeakyReLU-6&[-1, 100]&0\\
Linear-7&[-1, 4]&400\\
BatchNorm1d-8&[-1, 4]&8\\
LeakyReLU-9&[-1, 4]&0\\
Linear-10&[-1, 100]&400\\
BatchNorm1d-11&[-1, 100]&200\\
LeakyReLU-12&[-1, 100]&0\\
Linear-13&[-1, 4]&400\\
BatchNorm1d-14&[-1, 4]&8\\
LeakyReLU-15&[-1, 4]&0\\
Linear-16&[-1, 100]&400\\
BatchNorm1d-17&[-1, 100]&200\\
LeakyReLU-18&[-1, 100]&0\\
Linear-19&[-1, 6400]&640,000\\
BatchNorm1d-20&[-1, 6400]&12,800\\
LeakyReLU-21&[-1, 6400]&0\\
Linear-22&[-1, 1]&6,400\\
\end{tabular}
\end{ruledtabular}
\end{table}

\begin{table}[htb]
\caption{\label{discriminator_structure_tab}The structure of the discriminator for M-WGAN.}
\begin{ruledtabular}
\begin{tabular}{llr}
\textrm{Layer (type) }&
\textrm{Output Shape }&
\textrm{Parameter number }\\
\colrule
Linear-1&[-1, 64]&128\\
LeakyReLU-2&[-1, 64]&0\\
Linear-3&[-1, 3200]&208,000\\
LeakyReLU-4&[-1, 3200]&0\\
Linear-5&[-1, 1]&3,201\\
\end{tabular}
\end{ruledtabular}
\end{table}

The structure of the generator for M-WGAN is explained in Tab.~\ref{generator_structure_tab}. The fully connected layers are applied to reshape the input random latent variable with normal distribution. The batch normalization is exerted to improve generation. The structure of the discriminator for M-WGAN is described in Tab.~\ref{discriminator_structure_tab}. Three fully connected layers are used. LeakyReLU activation is implemented in all layers except for the output layer.

The input of generator for M-WGAN is a random latent variable tensor with shape [-1,1], and the output is a tensor with shape [-1,1], where -1 is an undetermined parameter. For example, we need to generate 1600 topological charge values, the input shape is [1600,1], and the output shape is also [1600,1]. Finally, the output needs to be reshaped into an 1600-dimensional vector. These 1600 data can form a distribution.

We update the parameters to maximize the Wasserstein distance and the optimizer is RMSProp\cite{arjovsky2017WGAN}. These settings are the same as those of the original WGAN.

As a result, the M-WGAN can generate the distribution of the topological charge directly to be applied to calculate the topological charge susceptibilities after training. The M-WGAN realize unsupervised generation without labels. In addition, some programs are based on Pytorch\cite{PASZKE2019PyTorch}.

The next part is the preparation of original data. The software Chroma\cite{Edwards_2005} is used to generate configurations of pure gauge field on individual workstation. In order to obtain the topological charge density data, the configurations are simulated first with the pseudo heat bath algorithm and smoothed by Wilson flow\cite{Gattringer2010}, then topological charge density data can be calculated from such configurations. In detail, the periodic boundary conditions and hot start have been applied and the updating steps are repeated 10 times for the visited link variable because the computation of sum of staples is costly. 

The Wilson flow step time $\varepsilon_f=0.01$ and total number of steps $N_{flow}=600$ are chosen.The reason why we use these two parameters are as follows.$\sqrt{8t_{f}}$ is used to characterize the smoothing range\cite{Luscher2010}\cite{vege2019nee}.In the section 6.2.1 of the literature\cite{vege2019nee}, the author said that when the flow smearing exceeds the lattice spacing, such that ${\sqrt{8t_{f}}}/a\gg1$, the discretization effects will be less visible.In this paper, we use the parameters $N_{flow} = 600$ and $\varepsilon_f=0.01$, so that ${\sqrt{8t_{f}}}/a\approx6.93\gg1$. Therefore, the degree of smearing is sufficient and $N_{flow}=600$ is enough to remove the Ultraviolet Divergences. In addition, referring to Appendix C of reference\cite{Luscher2010}, we obtain that the error of the Runge-Kutta scheme is small enough when $\varepsilon_f=0.01$.

\begin{figure}[htb]
\includegraphics[width=0.4\textwidth]{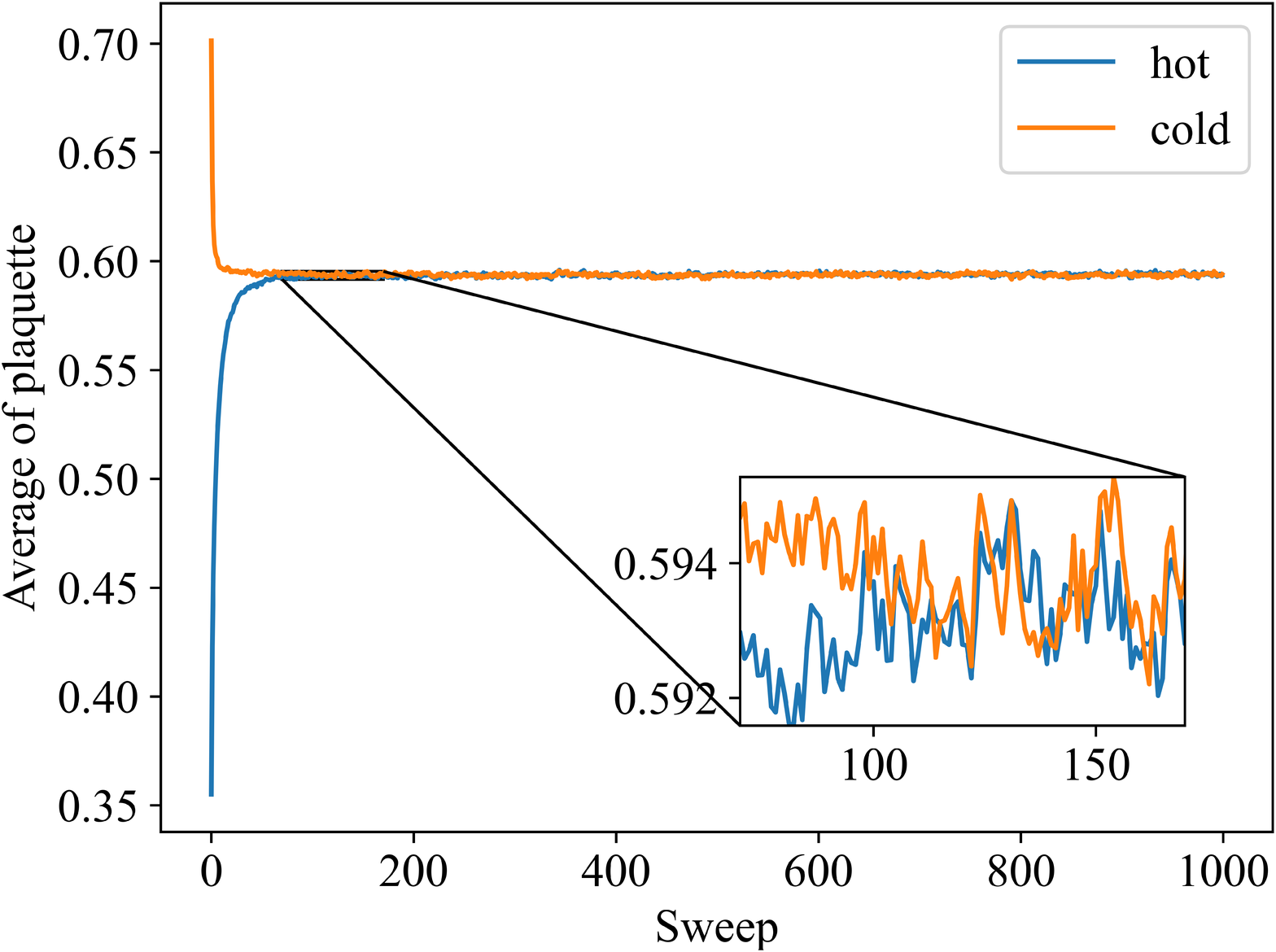}
\caption{\label{figure_2_ref}The evolution of the average of plaquette under different initial conditions.}
\end{figure}

The demonstration that the sampled configurations has reached thermalization is as follows. To detect whether the system has reached thermalization, it is possible to check a physical quantity starting from cold start and hot start is consistent after a certain number of MC steps. The average of the plaquette  with $\beta=6.0$ was chosen to study the thermalization and the mean value of plaquette is defined as $\left\langle {\frac{1}{3}ReTr{U_{plaq}}} \right\rangle$. It is found from Fig.~\ref{figure_2_ref}  that the system has reached thermalization after about 200 sweeps because the mean value of plaquette has evolved to a similar value starting from either cold or hot start..

Furthermore, it is necessary to calculate the integrated autocorrelation time which is computed by the topological charge in this article. For a Markov sequence of field configurations generated by MC, $X_i$ is a random variable and we could introduce the autocorrelation function
\begin{eqnarray}
{C_X}\left( {{X_i},{X_{i + t}}} \right) &&= \langle ({X_i} - \left\langle {{X_i}} \right\rangle )({X_{i + t}} - \left\langle {{X_{i + t}}} \right\rangle )\nonumber\\
 &&= \left\langle {{X_i}{X_{i + t}}} \right\rangle  - \left\langle {{X_i}} \right\rangle \left\langle {{X_{i + t}}} \right\rangle .
\end{eqnarray}
We define $C_X\left(t\right)\mathrm{=}C_X\left(X_i,\mathrm{\ } X_{i+t}\right)$ at thermalization and the normalized autocorrelation function $\Gamma_X\left(t\right)=\frac{C_X\left(t\right)}{C_X\left(0\right)}$. We introduce the integral autocorrelation time as 
\begin{equation}
{\tau _{X,int}}= \frac{1}{2} + \sum\limits_{t = 1}^N {{\Gamma _X}(t)} .
\end{equation}
It is calculated that the integrated autocorrelation time is 0.416. This result indicates that each of data can be regarded as independent because of $N/2\tau_{X,\mathrm{\ int\ }}>N$ \cite{Gattringer2010}, where a total of 1000 configurations are sampled with intervals of 200 sweeps.

In addition, the static QCD potential is applied to set the scale and can be parameterized by\cite{Gattringer2010} 
\begin{equation}
V\left(r\right)=A+\frac{B}{r}+\sigma r.
\end{equation}
The Sommer parameter $r_\mathrm{0}$ is defined as
\begin{equation}
\left(r^2\frac{dV\left(r\right)}{dr}\right)_{r=r_0}=1.65,
\end{equation}
and the Sommer parameter $\ r_\mathrm{0}\mathrm{\ =\ }0.49fm$ is used\cite{Sommer2014scale15}. The scales are
summarized in Tab.~\ref{static_potential_scale}.

\begin{table}[htb]
\caption{\label{static_potential_scale}The parameters used for the production of configurations and the associated value of $r_0$.}
\begin{ruledtabular}
\begin{tabular}{cccccc}
\textrm{$Volume$}&
\textrm{$\beta$}&
\textrm{$a[fm]$}&
\textrm{$N_{cnfg}$}&
\textrm{${r_0}/a$}&
\textrm{$L[fm]$}\\
\colrule
$24\times{12}^3$ & 6.0&0.093(3)&1000&5.30(15)&1.11(3) \\
\end{tabular}
\end{ruledtabular}
\end{table}

\section{Numerical results}
\begin{figure}[htb]
\includegraphics[width=0.5\textwidth]{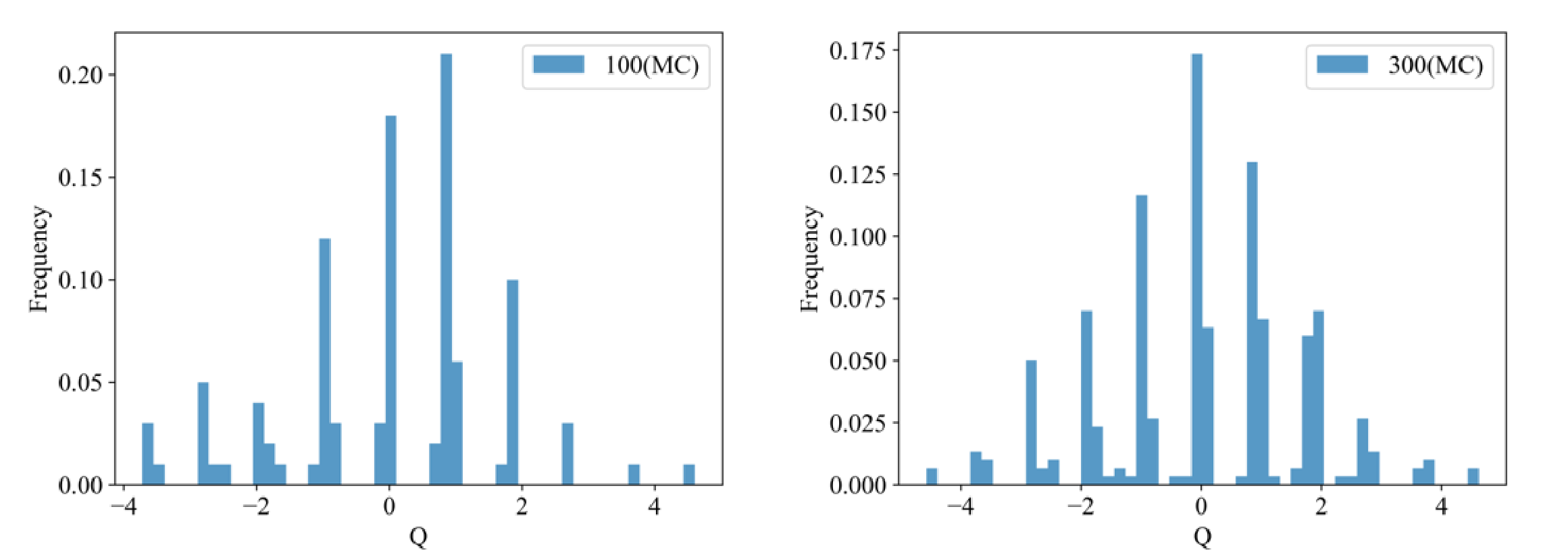}
\caption{\label{figure_3_ref}The distribution of topological charge based on the MC with Wilson flow. A total number of Wilson flow time steps $N_{flow}=600$ are chosen. The numbers of topological charge are 100 for the left subplot and 300 for the right subplot, respectively.}
\end{figure}

First, we show our results of 100 and 300 topological charge data simulated by the MC with Wilson flow, respectively. It is found from Fig.~\ref{figure_3_ref} that the topological charge $Q$ is approximately concentrated on the integer positions being consistent with the conclusion that the topological charge converges to an integer in the continuum limit mentioned above. It is calculated from the right subplot of Fig.~\ref{figure_3_ref} that the fourth root of topological susceptibility ${\chi_t}^{1/4}=191.8\pm3.9MeV$ when $N_{flow}=600$. Therefore, $a{\chi_t}^{1/4}$ is approximately equal to 0.09, , which is consistent with Refs\cite{Athenodorou_2020}. Moreover, the distribution of the topological charge should be symmetrical about the origin. However, we can find that the symmetry of distribution is poor in the left subgraph of Fig.~\ref{figure_3_ref} due to the poor statistics of data. Therefore, it is important to improve the distribution from the data of increased statistics. In MC simulations, increasing data may result in a rapid increase in computation time and storage usage, but these problems can be almost avoided with an appropriate ML model. Once the ML model is trained, it can immediately generate a corresponding data to improve the accuracy of the results. Next, we will discuss the details of two methods, the MC with Wilson flow and ML with the M-WGAN scheme, as well as combine two methods to generate configurations more efficiently.

For the MC method, we have simulated a total of 1600 configurations and calculated the topological charge from the configurations. Then the fourth root of the topological susceptibility ${\chi_t}^{1/4}=190.9\pm1.7MeV$ was calculated from 1600 topological charge data when 25 CPU cores were used in our calculations.

\begin{figure}[htb]
\includegraphics[width=0.5\textwidth]{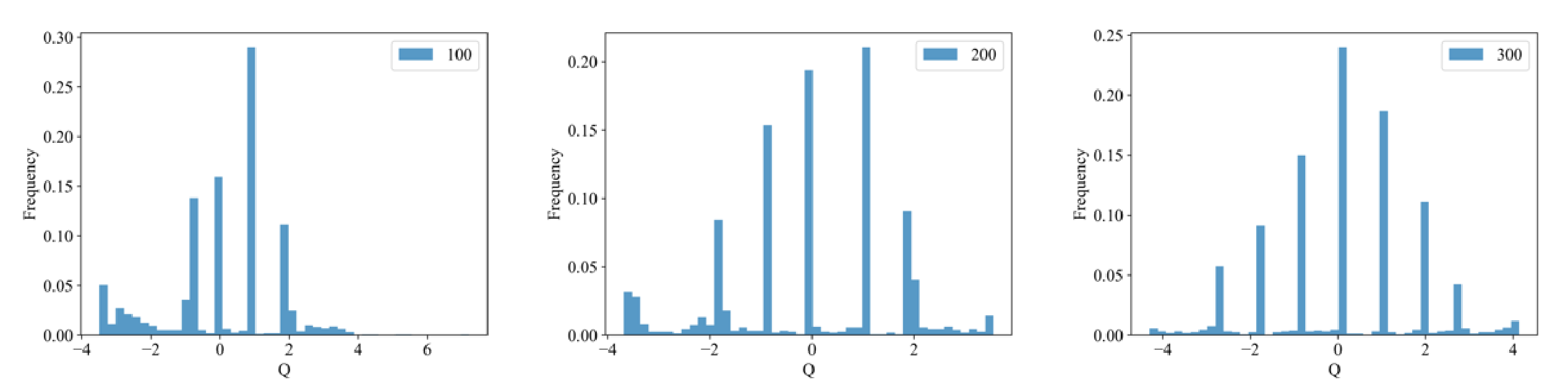}
\caption{\label{figure_4_ref}Comparison of models with 100, 200, 300 training data. }
\end{figure}

For ML, we concern to apply M-WGAN to generate the topological charge data based on the original MC training data. At first, we need to determine how much training data to be suitable. We tested the training processes with total training data volumes of 100, 200 and 300 and randomly select three-quarters of data volumes each iteration to train the model. The distribution of topological charge generated by the model trained using different data volumes is shown in the Fig.~\ref{figure_4_ref}. It can be observed that the distribution of the middle subgraph is mainly concentrated at the integers compared with the left subgraph, but its peak does not align with the origin. Furthermore, the distribution of the subgraph on the right is mainly concentrated at the integers and roughly symmetrical about the origin. Therefore, it is found that the model trains better as the amount of training data increases. By the test experiences, we choose to use 300 MC data as the ML training data sample and randomly pick out three-quarters of the 300 data in a training step to perform our ML train scheme. The total amount of training data we use is 300, and we randomly select three-quarters of 300 data, that is 225 data, to train the machine learning model each iteration. In other words, the batch size is 225. Compared with fixed selection, random selection can increase the diversity of data, which is more conducive to the training of machine learning model. In addition, three-quarters of 300 data are selected each time because such a large amount of data is more beneficial in training a better distribution.

\begin{figure}[htb]
\includegraphics[width=0.5\textwidth]{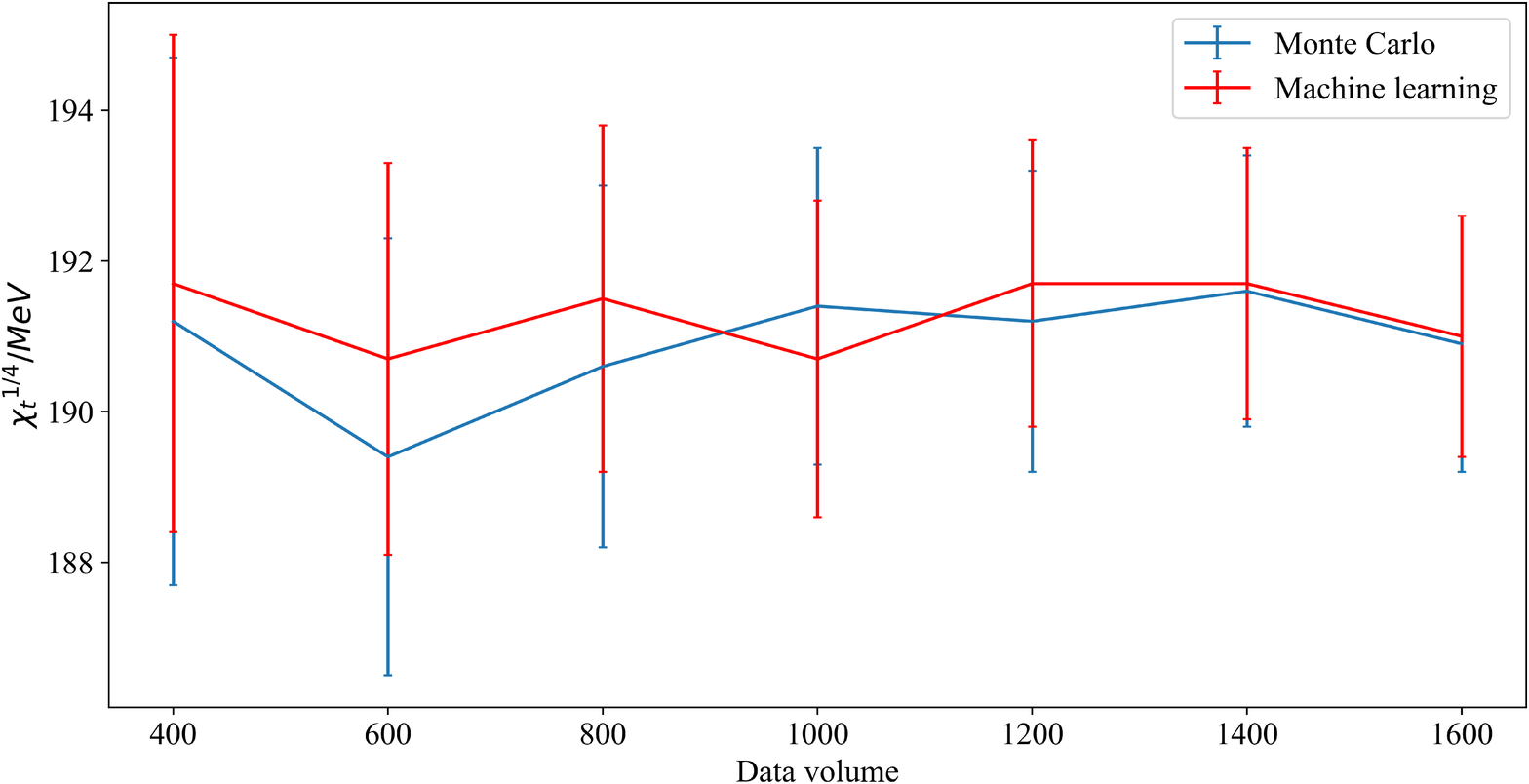}
\caption{\label{figure_5_ref}The fourth root of the topological susceptibility ${\chi_t}^{1/4}$ for MC and ML under different data size.}
\end{figure}

For our study,we should notice the accuracy of the physical results obtained with the MC and ML methods. It can be found from Tab.~\ref{MC and ML} and Fig.~\ref{figure_5_ref} that the data error gradually decreases as size of data increases for both MC and ML, where 300 training data were used to train M-WGAN and then 1600 output data were obtained. We apply the jack-knife to analysis the error of data in calculations\cite{Gattringer2010}.
\begin{table}[htb]
\caption{\label{MC and ML}The fourth root of the topological susceptibility ${\chi_t}^{1/4}$ for MC and ML.}
\begin{ruledtabular}
\begin{tabular}{ccc}
\textrm{Data volume}&
\textrm{${\chi_t}^{1/4}(MC)/MeV$}&
\textrm{${\chi_t}^{1/4}(ML)/MeV$}\\
\colrule
400 & $191.2\pm3.5$&$191.7\pm3.3$\\
600 & $189.4\pm2.9$&$190.7\pm2.6$\\
800 & $190.6\pm2.4$&$191.5\pm2.3$\\
1000 & $191.4\pm2.1$&$190.7\pm2.1$\\
1200 & $191.2\pm2.0$&$191.7\pm1.9$\\
1400 & $191.6\pm1.8$&$191.7\pm1.8$\\
1600 & $190.9\pm1.7$&$191.0\pm1.6$\\
\end{tabular}
\end{ruledtabular}
\end{table}

As shown in Tab.~\ref{time and storage}, the results from both methods are consistent within the statistical errors. But it is much faster to generate the same amount of data with the ML method, which also takes up less storage space than the MC method.  Therefore, we can apply ML to generate suitable data based on MC data to deal with the error more efficiently.
\begin{table}[htb]
\caption{\label{time and storage}The time consumption and hard disk storage usage for MC and ML. The time and storage of M-WGAN incorporate the effects of training data. }
\begin{ruledtabular}
\begin{tabular}{cccc}
\textrm{Method}&
\textrm{Time/h}&
\textrm{Storage/MB}&
\textrm{${\chi_t}^{1/4}/MeV$}\\
\colrule
MC & 136& 18230& $190.9\pm1.7$ \\
ML & 26& 3423& $191.0\pm1.6$\\
\end{tabular}
\end{ruledtabular}
\end{table}

\begin{figure}[htb]
\includegraphics[width=0.5\textwidth]{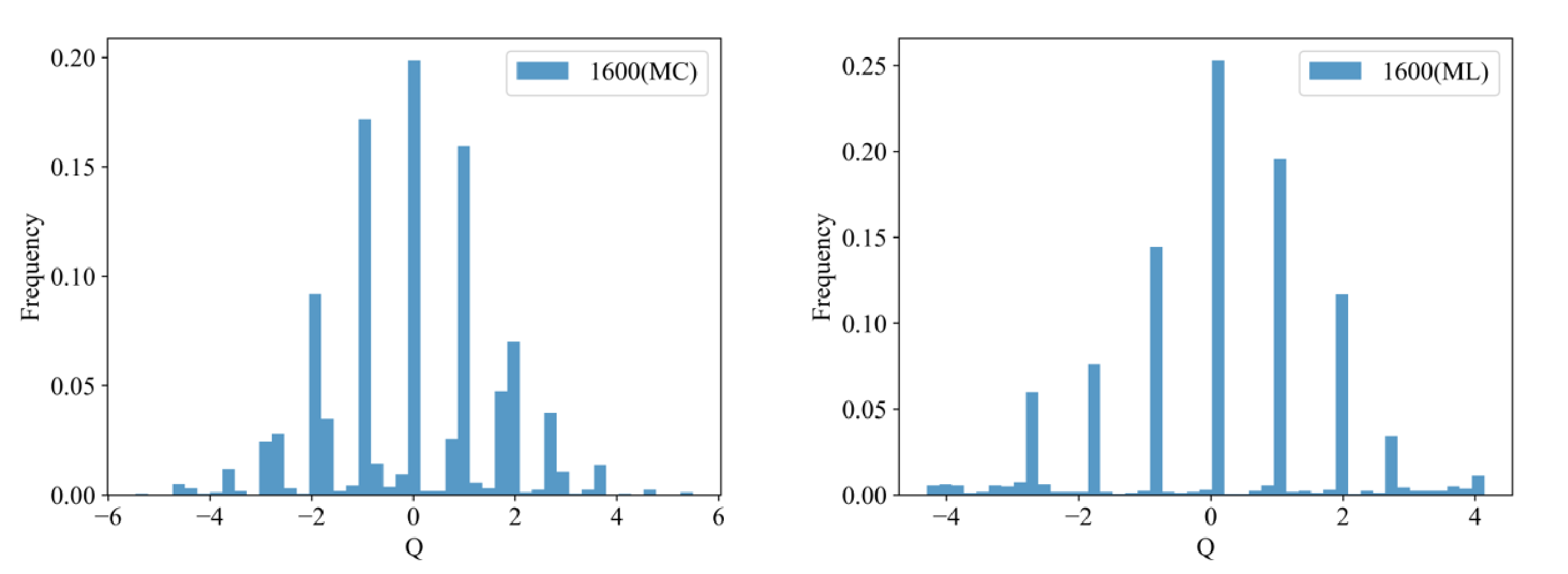}
\caption{\label{figure_6_ref}The distributions of topological charge for MC and ML.}
\end{figure}

Furthermore, the distributions of topological charge generated by the MC and ML are shown in Fig.~\ref{figure_6_ref}. We can conclude that both the distributions have good symmetry, and their data are discretely distributed at integers. These characteristics are consistent with the nature of the topological charge. In addition, the integrated autocorrelation time of 1600 data for ML is 0.41, which indicates that these data are independent.

\section{Conclusion}
In this paper, we have studied the topological quan-tities of lattice QCD by using MC simulation combined with the M-WGAN . Especially we have applied the M-WGAN to generate the distribution of topological charge to show the potential for applications of ML technique in a study of lattice QCD.

By our experience, the conclusions are as follows. Compared with the MC by the pseudo heat bath algorithm and the Wilson flow, the M-WGAN shows its efficiency for the MC simulations in terms of time cost and storage as shown in Tab.~\ref{MC and ML} and Tab.~\ref{time and storage}. The data generated by the M-WGAN trained with 300 data are more accurate than the corresponding data by the MC simulation as the topological susceptibility is concerned. The pseudo distribution generated by this model after a necessary training can be applied to calculate the correct topological susceptibility in the SU(3) lattice QCD.

For future studies, we hope that the M-WGAN can be applied to tackle other physics problems in lattice QCD and provide an alternative approach to simulating some interesting and important quantities in a much more eﬃcient way.

\textbf{Acknowledgments.}  This work was done based on the Chroma applied to simulate the conﬁgurations of the lattice gauge ﬁeld. We are grateful to the relevant contributors to the Chroma, and Heping Ying would like to thank all his co-workers on the career for lattice QCD field studies.

\bibliographystyle{apsrev4-2}
\bibliography{apssamp}

\end{document}